\documentstyle[aps,epsf,bezier]{revtex}

\newcommand{\bqn}{\begin{eqnarray}}
\newcommand{\eqn}{\end{eqnarray}}
\newcommand{\beq}{\begin{equation}}
\newcommand{\eeq}{\end{equation}}
\newcommand{\Hh}{{\textstyle\frac 12}}

\newcommand{\mbf}[1]{\mbox{\boldmath$#1$}}

\def\bk{{\mbox{\boldmath$k$}}}
\def\bq{{\mbox{\boldmath$q$}}}

\def\bke{{\mbox{\boldmath$k$}_{e}}}

\def\bp{{\mbox{\boldmath$p$}}}
\def\br{{\mbox{\boldmath$r$}}}
\def\bP{{\mbox{\boldmath$P$}}}
\def\bPp{{\mbox{\boldmath$P$}^{\prime}}}

\newcommand{\CM}{{\cal M}}

\def\Sp{^1S_0^{+}}
\def\Sm{^1S_0^{-}}

\def\PAe{^3P_0^{e}}
\def\PAo{^3P_0^{o}}

\begin{document}

\title{Deuteron electrodisintegration near threshold in the
  Bethe--Salpeter approach}

\author{ S.G. Bondarenko, V.V. Burov} \address{Bogoliubov Laboratory
  of Theoretical Physics, JINR Dubna, 141980 Russia} \author{M. Beyer}
\address{Max Planck AG ``Theoretical Many Particle Physics'', Rostock
  University, 18051 Rostock, Germany} \author{S.M. Dorkin}
\address{Far Eastern State University Vladivostock, 690000 Russia}

\maketitle

\begin{abstract}
  In the framework of the covariant Bethe--Salpeter formalism we
  calculate the amplitude of electrodisintegration of the deuteron
  near the threshold energy.  Due to the small relative momentum of
  the final $np$-pair, the deuteron break--up process may be
  sufficiently described by a transition to the dominant $^1S_0$
  channel into the final state.  Therefore only one unknown structure
  function (form factor) exists.  The explicit formula for this form
  factor is given in terms of partial Bethe--Salpeter amplitudes.  We
  compare our result to the usual non--relativistic calculations.  The
  essential role of $P$--states as well as their possible link to
  mesonic exchange currents is discussed.
\end{abstract}


\section{Introduction}

Disintegration of the deuteron (either by photons or electrons) has
been and still is a rich source of information on the structure of
electromagnetic current operators and nuclear dynamics. In the context
of photodisintegration a recent overview on experimental issues and
theoretical methods is given in ref.~\cite{arsanz}.

Here we touch upon the problem of deuteron electrodisintegration near
the threshold energy. Due to the small value of the relative momentum
($p^*$, in the rest system) of the final $np$-pair it is reasonable to
suppose that the $^1S_0$ channel dominates in the final
state~\cite{mathiot}. So the experiment is well suited to study the
deuteron wave function in a pure way.  Recent experiments have been
done in Saclay for $E_{np}=0-3$ MeV~\cite{sacle}, and at SLAC with
$E_{np}=0-10$ MeV~\cite{slac} ($E_{np}$ is the relative energy of the
outcoming neuteron proton pair).

The disintegration process is usually described within a
non--relativistic framework including relativistic corrections.  Only
one investigation is known to have used a relativistic description of
the process considered here in the framework of light front
dynamics~\cite{karmanov}.

It is recognized, that the non--relativistic impulse approximation
does not reproduce the experimental differential cross section even
for small momentum transfers. Actually, the minimum that appears in
the nonrelativistic impulse approximation is not present in the
experimental data~\cite{sacle,slac,mathiot}. This minimum, however, is
filled with contributions of mesonic exchange currents (MEC).  In its
turn, this filling is a strong confirmation of MEC. Among the MEC
considered, the most important ones are related to the pion
exchange~\cite{mathiot,tamura}. In addition, $\rho$--meson exchange
currents, $\Delta$--isobar configurations, and final state
interactions as well as retardation effects, are to be taking into
account~\cite{mathiot,tamura,burov}.

Generally, the differential cross section is very sensitive to the
choice of the neutron form factor, vertex form factors, and the way of
constructing mesonic exchange currents. The latter strongly depends on
the form of the $\gamma NN$ form factor ($F_1$, or $G_E$) that
influences the results of the cross section~\cite{f1ge,Geschool}.

The main goal of the present work is to describe the deuteron
electrodisintegration in a consistent relativistic approach. This will
be done with the Bethe--Salpeter equation for the two nucleon system.
This way is possible to come to general conclusions about the
amplitude of the process not seen in the non--relativistic approach.
On the other hand, the non--relativistic limit will be recovered and
some links to the non--relativistic corrections can be established.

In addition, we consider polarization observables in the covariant
approach and give some explicit examples for polarized electrons on
polarized deuteron scattering. The non--relativistic formalism may be
found in ref.~\cite{alt93} and the helicity approach of polarization
in ref.~\cite{gross89}.

The paper is organized as follows. In the next section the general
form of the transition amplitude for deuteron break--up near the
threshold energy is introduced and differential cross-sections for
unpolarized and polarized electrons and deuterons are given.  In
section III relations between the asymmetries are deduced. Section IV
is devoted to the covariant formalism of the Bethe--Salpeter approach,
and in section V we evaluate the amplitude in the general form.
Different approximations are considered in section VI. The connections
to nonrelativistic calculations are discussed in section VII. The last
section contains our conclusions and outlook.

\section{The deuteron electrodisintegration near threshold cross
  section: general formulae} 

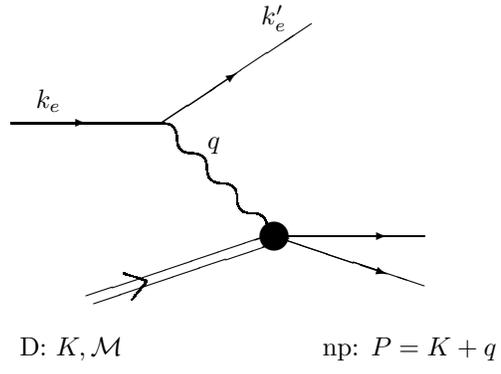
\begin{figure}[h]

\unitlength=1mm
\special{em:linewidth 0.4pt}
\linethickness{0.4pt}

\hskip 30mm
\begin{picture}(50.00,70.00)
\put(20.00,60.00){\line(1,0){20.00}}
\put(40.00,60.00){\line(3,2){20.00}}
\put(20.00,60.00){\vector(1,0){10}}
\put(40.00,60.00){\vector(3,2){10}}
\put(55.00,45.00){\line(1,0){20.00}}
\put(55.00,45.00){\vector(1,0){15.00}}
\put(55.00,45.00){\line(3,-1){20.00}}
\put(55.00,45.00){\vector(3,-1){15.00}}
\put(30.00,37.00){\line(3,1){26.00}}
\put(31.00,36.00){\line(3,1){26.00}}
\bezier{40}(35.00,40.00)(38.00,39.00)(38.00,39.00)
\bezier{40}(36.00,36.50)(38.00,39.00)(38.00,39.00)
\put(55.00,45.00){\circle*{4.00}}
\bezier{40}(40.00,60.00)(42.00,60.00)(42.00,58.00)
\bezier{40}(42.00,58.00)(42.00,56.00)(44.00,56.00)
\bezier{40}(44.00,56.00)(46.00,56.00)(46.00,54.00)
\bezier{40}(46.00,54.00)(46.00,52.00)(48.00,52.00)
\bezier{40}(48.00,52.00)(50.00,52.00)(50.00,50.00)
\bezier{40}(50.00,50.00)(50.00,48.00)(52.00,48.00)
\bezier{40}(52.00,48.00)(54.00,48.00)(54.00,46.00)
\put(28.00,30.00){\makebox(0,0)[cc]{D: $K,{\cal M}$}}
\put(73.00,30.00){\makebox(0,0)[cc]{np: $P=K+q$}}
\put(47.00,57.00){\makebox(0,0)[cc]{$q$}}
\put(25.00,63.00){\makebox(0,0)[cc]{$k_e$}}
\put(55.00,74.00){\makebox(0,0)[cc]{$k_e^{\prime}$}}
\end{picture}

\vskip -20mm
\caption{The one photon approximation.}
\label{fig:one}
\end{figure}

The general form of the deuteron electrodisintegration amplitude
$M_{fi}$ in the one photon approximation (Fig.~\ref{fig:one}) may be
written in the following way~\cite{bjorken},
\begin{eqnarray}
M_{fi}\;=\;ie^2\;{\bar u(k_e^{\prime},s_e^{\prime})}\;
\gamma^{\mu}u(k_e,s_e)\;\frac{1}{q^2}
\;\langle np | j_{\mu} | D {\cal M} \rangle,
\label{eqn:M}
\end{eqnarray}
where $u(k_e,s_e)$ denotes the free electron spinor with 4-momentum
$k_e$ and spin $s_e$, and $q=k_e-k_e^{\prime}$ is the 4-momentum
transfer.  The hadronic transition matrix element $\langle np |
j_{\mu} | D {\cal M} \rangle$ leads from the deuteron state $| D {\cal
  M} \rangle$ with 4-momentum $K$ and total angular momentum
projection $\cal M$ to the final $np$ state with 4-momentum $P=K+q$,
where $j_\mu$ is the electromagnetic current operator.

Near the threshold energy the final $np$ state may be assumed to be
dominantly in the $^1S_0$ state~\cite{dominant} and the hadronic
matrix element may be considered then as a $1^+\rightarrow 0^+$
transition. Due to the Lorentz transformation properties and because
of parity and time reversal symmetry, the general form is restricted
to only one contribution, i.e.
\begin{eqnarray}
\langle np (^1S_0) | j_{\mu} | D {\cal M} \rangle =
i\epsilon_{\mu\alpha\beta\gamma}\;\xi^{\alpha}_{\cal M}\;
q^{\beta}\;K^{\gamma}\;V(s,q^2)\;
\equiv\;\xi^{\alpha}_{\cal M}\;G_{\mu\alpha}\;V(s,q^2),
\label{formc}
\end{eqnarray}
where $\xi^{\alpha}_{\cal M}$ is the polarization 4-vector of the
deuteron, and $V(s,q^2)$ a Lorentz scalar real function of invariants
$q^2$ and $s=P^2 \approx 4m^2$. Below this structure will be shown
explicity by straightforward calculations. Note that due to the
antisymmetric tensor $\epsilon_{\mu\alpha\beta\gamma}$ this transition
current meets the requirement of gauge independence i.e.  $q_{\mu}
\langle np (^1S_0) | j_{\mu} | D {\cal M} \rangle=0$.  Thus the
evaluation of the amplitude in the framework of the Bethe --Salpeter
approach in the form of eq. (\ref{formc}) will prove the non trivial
property of gauge independence.

Using eq.~(\ref{eqn:M}) the differential cross-section may be obtained
in the standard way, see ref.~\cite{bjorken}.  Here it is useful to
introduce leptonic $l_{\mu\nu}$ and hadronic $W^{\mu\nu}$ tensors.
Then the differential cross section reads as
\begin{eqnarray}
\frac{d^{2}\sigma}{dE_e^{\prime}d\Omega_e^{\prime}}\; =\;
\frac{\alpha^2}{q^4}\; \frac{|\bke^{\prime}|}{|\bke|}\;
l_{\mu\nu}\; W^{\mu\nu},
\end{eqnarray}
with $\alpha=e^2/4\pi$.  The leptonic tensor is given by
\begin{equation}
l_{\mu\nu} = 2 (k_{e\mu}k^{\prime}_{e\nu}+k^{\prime}_{e\mu}k_{e\nu})+
q^2 g_{\mu\nu}
+2im_{e}{\epsilon}_{\mu\nu\alpha\beta}q^{\alpha}s_{e}^{\beta}.
\end{equation}
The hadronic tensor $W_{\mu\nu}$ has the following form:
\begin{eqnarray}
W^{\mu\nu} =
\langle np (^1S_0) | j^{\mu} | D {\cal M} \rangle
\langle D {\cal M} | j^{\dagger\nu} | np (^1S_0)\rangle
\label{W:def}
\frac{(2\pi)^3}{2M}
\int \delta(K+q-k_p-k_n) \frac{d{\bk}_p}{2E_{p}(2\pi)^3}
\frac{d{\bk}_n}{2E_{n}(2\pi)^3}.
\nonumber
\end{eqnarray}
Using the general form of the hadronic transition current
eq.(\ref{formc}), the hadronic tensor may be written as
\begin{equation}
W^{\mu\nu} = R\;G^{\mu\alpha}\;\rho_{\alpha\beta}\;G^{*\nu\beta}\;
\Bigl|V(s,q^2)\Bigr|^2,
\label{eqn:W2}
\end{equation}
where $R$ is a purely kinematical factor. It is given by
\begin{equation}
R=\frac{1}{8\pi^2}\frac{1}{2M} \frac{|\bp^{*}|}{\sqrt{s}},
\qquad |\bp^{*}|=\sqrt{\frac{s}{4}-m^2}.
\end{equation}
Here $m$ is the nucleon mass and $M$ is the deuteron mass.

In eq.~(\ref{eqn:W2}) the density matrix $\rho_{\alpha\beta}$ of the
deuteron reads
\begin{eqnarray}
\rho_{\alpha\beta} = \frac{1}{3}(-g_{\alpha\beta}+
\frac{K_{\alpha}K_{\beta}}{M^2})
+\frac{1}{2M}i
{\epsilon}_{\alpha\beta\gamma\delta} K^{\gamma} s_{D}^{\delta}
&& -\bigl[\frac{1}{2}
\bigl(
{(W_{\lambda_1})}_{\alpha\rho}
{(W_{\lambda_2})}^{\rho}_{~\beta}+
{(W_{\lambda_2})}_{\alpha\rho}
{(W_{\lambda_1})}^{\rho}_{~\beta}
\bigr)
\\
&& +\frac{2}{3}
(-g_{\lambda_1\lambda_2}+\frac{K_{\lambda_1}K_{\lambda_2}}{M^2})
(-g_{\alpha\beta}+\frac{K_{\alpha}K_{\beta}}{M^2}) \bigr] p_{D}^{\lambda_1
\lambda_2},
\nonumber\end{eqnarray}
where
${(W_{\lambda})}_{\alpha\beta}=
i\epsilon_{\alpha\beta\gamma\lambda}K^{\gamma}/M$,
$s_{D}$ is the spin vector and $p_{D}$ is the alignment tensor
of the deuteron.  Using this explicit form of the density
matrix the hadronic tensor may then be written as
(electron mass is neglected)
\begin{eqnarray}
&& W^{(u)}_{\mu\nu}=\frac{1}{3}\;R\;
\bigl[g_{\mu\nu}(q^2M^2-(Kq)^2)
+(K_{\mu}q_{\nu}+q_{\mu}K_{\nu})(Kq)
-K_{\mu}K_{\nu}q^2-q_{\mu}q_{\nu}M^2\bigr]\;\Bigl|V(s,q^2)\Bigr|^2,
\nonumber\\
&& W^{(v)}_{\mu\nu}=\frac{1}{2}\;R\;M\;(s_Dq)\;
i {\epsilon}_{\mu\nu\alpha\beta}\;q^{\alpha}\;K^{\beta}\;
\Bigl|V(s,q^2)\Bigr|^2,
\label{tensor}\\
&& W^{(t)}_{\mu\nu}=R\;\Bigl[\frac{1}{2}[\epsilon_{\mu\lambda_2\alpha\beta}
\epsilon_{\lambda_2\nu\gamma\delta}+\epsilon_{\mu\lambda_2\alpha\beta}
\epsilon_{\lambda_1\nu\gamma\delta}]K^{\alpha}K^{\gamma}q^{\beta}q^{\delta}
+\frac{1}{3}\bigl(-g_{\lambda_1\lambda_2}+\frac{K_{\lambda_1}K_{\lambda_2}}
{M^2}\bigr)\bigl[g_{\mu\nu}(q^2M^2-(Kq)^2)
\nonumber\\
&& \hskip 20mm +(K_{\mu}q_{\nu}+q_{\mu}K_{\nu})(Kq)-K_{\mu}K_{\nu}q^2
-q_{\mu}q_{\nu}M^2\bigr]\;p_D^{\lambda_1\lambda_2}
\Bigr]\;\Bigl|V(s,q^2)\Bigr|^2.
\nonumber\end{eqnarray}
The superscripts ${(u,v,t)}$ denote unpolarized, vector polarized and
tensor polarized cases, respectively.

Thus for the case of unpolarized electrons and deuterons the
differential cross section can be written as
\begin{eqnarray}
\frac{d^2\sigma}{dE^{\prime}_e d \Omega^{\prime}_e}=
\Bigl(\frac {d \sigma}{d \Omega}\Bigr)_M\;
\frac{M\,|\bp^{*}|\,m}{12\pi^2}\;
[(E_e+E_e^{\prime})^2-2E_eE_e^{\prime}\cos^2{\frac{\theta_e}{2}}]\;
\tan^2{\frac{\theta_e}{2}}\; \Bigl|V(s,q^2)\Bigr|^2,
\label{main}\end{eqnarray}
where 
\begin{equation}
\Bigl(\frac {d \sigma}{d \Omega}\Bigr)_M = \frac{\alpha^2\, 
\cos^2{\frac{\theta_e}{2}}}{4\,E^2\,\sin^4{\frac{\theta_e}{2}}}
\end{equation}
is the Mott cross section. The normalization conditions of state
vectors have been chosen to be for deuteron
$$<D\bPp \CM^{\prime}|D\bP \CM>=2E_D\,(2\pi)^3\,\delta_{\CM \CM^{\prime}}
\delta(\bPp-\bP)$$ 
and for nucleons 
$$<N\bp,\mu |N \bk,\mu^{\prime}> =
2E_N\,(2\pi)^3\,\delta_{\mu \mu^{\prime}}\delta(\bp-\bk)$$.

\section{Asymmetries for polarized deuterons}

\begin{figure}[h]

\unitlength=1mm
\special{em:linewidth 0.4pt}
\linethickness{0.4pt}

\hskip 20mm
\begin{picture}(100.00,90.00)
\put(48.50,90.00){\line(1,0){85.00}}
\put(100.00,40.00){\line(-1,0){85.00}}
\put(133.50,90.00){\line(-2,-3){33.40}}
\put(15.00,40.00){\line(2,3){33.40}}
\put(40.00,60.00){\vector(1,0){30.00}}
\put(70.00,60.00){\vector(4,3){20.00}}
\put(70.00,60.00){\line(1,0){22.00}}
\put(95.00,55.00){\vector(1,0){10.00}}
\put(95.00,55.00){\vector(-3,-4){6.00}}
\put(95.00,55.00){\vector(3,1){10.00}}
\bezier{32}(75.00,63.50)(76.00,62.00)(76.00,60.00)
\put(81.00,63.00){\makebox(0,0)[cc]{$\theta_e$}}
\put(52.00,63.00){\makebox(0,0)[cc]{$\bk_e$}}
\put(76.00,70.00){\makebox(0,0)[cc]{$\bk_e^{\prime}$}}
\put(86.00,45.00){\makebox(0,0)[cc]{$X$}}
\put(104.00,61.00){\makebox(0,0)[cc]{$Y$}}
\put(103.00,52.00){\makebox(0,0)[cc]{$Z$}}
\end{picture}

\vskip -30mm
\caption{The kinematics for $ed\to enp$.}
\label{fig:kinem}
\end{figure}
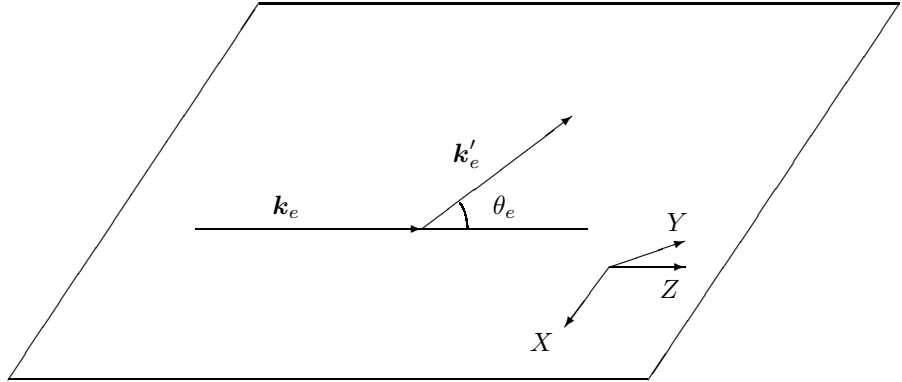

With the general form of hadronic tensor $W^{\mu\nu}$
eq.~(\ref{tensor}), it is straight forward to calculate asymmetries of
the deuteron disintegration near threshold, e.g.
\begin{eqnarray}
A=\frac {d\sigma(\uparrow,D)-d\sigma(\downarrow,D)}
 {d\sigma(\uparrow,D)+d\sigma(\downarrow,D)},
\label{assym}
\end{eqnarray}
where $d\sigma$ is the differential cross section,
$\uparrow(\downarrow)$ denotes the helicity $\lambda_e=+1(-1)$ of the
incoming electron and $D$ the polarization state of the deuteron. We
assume the initial electron moving along the $Z$-axis, and $\theta_e$
is the electron scattering angle. The scattering plane of the electron
is in the $XZ$ plane (see Fig~\ref{fig:kinem}).  Then the vectors
$k_e$ and $k_e^{\prime}$ are of the form:
\begin{eqnarray}
  k_e=(E_e,0,0,E_e), \quad\quad
  k_e^{\prime}=
(E_e^{\prime},-E_e^{\prime}\sin{\theta_e},0,E_e^{\prime}\cos{\theta_e}).
\end{eqnarray}

First we consider the case of {\em vector} polarized deuterons. If the
direction of the deuteron polarization is parallel to the $Z$-axis,
then
\begin{eqnarray}
A_{\parallel} &=& \frac{3}{2}\ \kappa
\frac{(E_e+E_e^{\prime})
(E_e-E_e^{\prime}\cos{\theta_e})}{(E_e+E_e^{\prime})^2
-2E_eE_e^{\prime}\cos^{2}{\theta_e/2}},
\label{A:parallel}\end{eqnarray}
where $\kappa$ is the degree of polarization of the deuterons. Note,
that the dependence on the form factor $V(s,q^2)$ disappears.  For the
case of the backward scattering ($\theta_e=180^{\circ}$)
eq.~(\ref{A:parallel}) even simplifies to
\begin{equation}
A_{\parallel} = \frac{3}{2}\ \kappa.
\end{equation}
If the polarization of the deuteron is parallel to the $X$-axis, then
\begin{eqnarray}
A_{\perp} = \frac{3}{2}\ \kappa\ \frac{(E_e+E_e^{\prime})
E_e^{\prime}\sin{\theta_e}}
{(E_e+E_e^{\prime})^2-2E_eE_e^{\prime}\cos^{2}{\theta_e/2}}.
\label{A:perp}
\end{eqnarray}

These formulae may be generalized to arbitrary polarization direction
of the deuteron  given by the angles $(\vartheta,\varphi)$, viz.
\begin{eqnarray}
A(\vartheta,\varphi) = \frac{3}{2}\ \kappa
\frac{(E_e+E_e^{\prime})(E_e^{\prime}\sin{\theta_e}
\sin{\vartheta}\cos{\varphi}+
(E_e-E_e^{\prime}\cos{\theta_e})\cos{\vartheta})}
{(E_e+E_e^{\prime})^2-2E_eE_e^{\prime}\cos^{2}{\theta_e/2}}.
\label{A:arb}\end{eqnarray}

Now, consider the case of {\em tensor} polarization of the inital
target.  If the initial deuteron is only aligned due to a $p_{D\,zz}$
component, then the cross section reads
\begin{eqnarray}
&&d\sigma(p_{D\,zz})=d\sigma[1 + A_{zz} p_{D\,zz}],
\label{ratio}\\
&& A_{zz}=\frac{4E_e^2+E_e^{\prime\;2}
+4E_eE_e^{\prime}-4E_eE_e^{\prime}\cos{\theta_e}+
3E_e^{\prime\;2}\cos{2\theta_e}}
{4((E_e+E_e^{\prime})^2-2E_eE_e^{\prime}\cos^2{\theta_e/2})},
\nonumber\end{eqnarray}
where $A_{zz}$ is the tensor analyzing power. 
For the backward scattering the analyzing power is simplifies
\begin{eqnarray}
A_{zz} = 1.
\label{rback}\end{eqnarray}

\section{Initial and final states in the covariant formalism}

The main object of the Bethe--Salpeter approach is the Bethe--Salpeter
amplitude (or wave function), which is usually decomposed into a sum
of direct products of free Dirac spinors. In order to achieve such a
decomposition, an additional quantum number for the moving nucleon has
been introduced besides spin and isospin, which is the $\rho$--spin.
The operators of spin $S_i$ and $\rho$-spin $R_i$ ($i=1,2,3$) may be
represented in the following way:
\begin{eqnarray}
R_1&=&\gamma_5({\bf k} \cdot\mbf{\gamma}+m)/E_k,
\nonumber\\
R_2&=&-i\gamma_0 \gamma_5,
\nonumber\\
R_3&=&\gamma_0 ({\bf k} \cdot \mbf{\gamma} +m), 
  \label{eqn:R}
\end{eqnarray}
and
\begin{equation}
S_i=-\gamma_0 \gamma_5 \bigr(\frac{m}{E_k} \gamma_i - \frac {k_i}{E_k} -
\frac{{\bf k}\cdot \mbf{\gamma} } {E_k} \gamma_i 
+\frac {{\bf k}\cdot \mbf{\gamma}
  k_i} {E_k(m+E_k)}\bigl),
  \label{eqn:S}
\end{equation}
where ${\bf k}$ is the momentum of the nucleon and
$E_k=\sqrt{k^2+m^2}$.  The commutation relations for these operators
are: \bqn ~[S_i,S_j]&=&2i\epsilon_{ijk} S_k, \quad [S^2,S_i]=0,
\nonumber\\
~[R_i,R_j]&=&2i\epsilon_{ijk} R_k, \quad [R^2,R_i]=0,
\nonumber\\
~[S_i,R_j]&=&0, \nonumber \eqn where $S^2=\sum_i S^2_i$ and
$R^2=\sum_i R^2_i$.  In the case of ${\bf k}=0$ these operators have a
simple form of a direct product of matrices acting independently in
$\rho$ and $\sigma$ spaces, viz.
\begin{equation}
S_i=I_{\rho}\otimes \sigma_i\qquad
R_i=\rho_i\otimes I_{\sigma},
\nonumber
\end{equation}
where $\sigma_i (\rho_i)$ are Pauli matrices and 
$I_{\sigma}(I_{\rho})$ are unit matrices in the spin ($\rho$--spin) space. The
eigenvector of operators $S^2,S_3,R^2,R_3$ is denoted as
$U^{\rho}_\mu({\bf k})$ with $\rho$ projection of $\rho$--spin and
$\mu$ of the usual spin. Then using 
\bqn 
 S_3 U^{\rho}_{\mu}({\bf
  k})=\mu U^{\rho}_{\mu}({\bf k}), &\quad &S^2 U^{\rho}_{\mu}({\bf k})
=
3U^{\rho}_{\mu}({\bf k}), \\
R_3 U^{\rho}_{\mu}({\bf k})=\rho U^{\rho}_{\mu}({\bf k}), &\quad &R^2
U^{\rho}_{\mu}({\bf k}) =3U^{\rho}_{\mu}({\bf k}), 
\eqn 
one obtains
\bqn 
U^{\rho=+1}_{\mu} ({\bf k})=u_{\mu}({\bf k}), \quad
U^{\rho=-1}_{\mu} ({\bf k})=v_{\mu}(-{\bf k}) , \nonumber 
\eqn 
where
$u_{\mu}({\bf k})$ and $v_{\mu}({\bf k})$ are the free Dirac spinors
and spin index $\mu=\pm1$ corresponds to the spin projection $\pm
\frac12$.

For the Dirac spinors it is convenient to use the following notation:
\begin{eqnarray}
u_{\mu}({\bf k}) =
\hat {\cal L}({\bf k}) u_{\mu} ({\bf 0}), \quad
v_{\mu}({\bf k}) =
\hat {\cal L}({\bf k}) v_{\mu} ({\bf 0}).
\label{trans}
\end{eqnarray}
where the boost of particle with mass $m$ and spin-$\Hh$ is given by
\begin{eqnarray}
{\cal L}({\bf k}) =
 \frac {m+\hat k_1 \gamma_0}{\sqrt{2E_k(m+E_k)}},
\label{lor}
\end{eqnarray}
with the notation $\hat k = k_{\mu}\gamma^\mu$,
and  the nucleon 4-momentum
$k_1=(E_k,{\bf k})$.
In the rest frame of the particles the spinors are given by
\begin{eqnarray}
u_{\mu}({\bf 0}) =
\left(\begin{array}{c} \chi_{\mu}  \\ 0 \end{array} \right),  \quad
v_{\mu}({\bf 0})=
\left(\begin{array}{c}         0   \\ \chi_{\mu} \end{array} \right).
\nonumber
\end{eqnarray}
Another useful formula is 
\begin{eqnarray}
u_{\mu}(-\bk) =
 \frac {m+\hat k_2 \gamma_0}{\sqrt{2E_k(m+E_k)}} u_{\mu}(\bf 0),
\label{ror}
\end{eqnarray}
with $k_2=(E_k,-\bk)$.  The Bethe -- Salpeter wave function of the
two-nucleon state with angular momentum $J$ and its projection ${\cal
  M}$ in the rest frame is written as follows (see also
~\cite{honzava}):
\begin{eqnarray}
\Phi_{J{\cal M}}(K,k)=\sum \limits_{\alpha} \;
g_{\alpha} (k_0,|\bk|)\;
{\cal Y}_{\cal M}^{\alpha}({\bf k}),
\label{reldp}
\end{eqnarray}
where $K$ is the total momentum of the system (in the rest frame
$K=K^{(0)}=(M,{\bf 0})$), and $k=(k_0,\bk)$ is the relative momentum.
The decomposition is according to the quantum numbers of relative
orbital angular momentum $L$, total spin $S$, and $\rho$-spin
collectively denoted by $\alpha$ \cite{kubis}. The radial parts of the
wave function are denoted by $g_{\alpha} (k_0,|\bk|)$, and the
spin-angular momentum part ${\cal Y}_{\cal M}^{\alpha}({\bf k})$ is
given by
\begin{eqnarray}
{\cal Y}_{\cal M}^{\alpha}({\bf k})&=&
i^L \sum \limits_{\mu_1,  \mu_2 m_L,\rho_1,\rho_2}\;
(L m_L S m_S | J {\cal M}) \;
(\Hh \mu_1 \Hh \mu_2 | S m_S)\;
 \nonumber\\ &&\times 
(\Hh \rho_1 \Hh \rho_2 | \rho m_{\rho})\;
Y_{L m_L}(\Omega_{\bk})\;
{U_{\mu_1}^{\rho_1}}^{(1)}({\bf k})\;
{U_{\mu_2}^{\rho_2}}^{(2)}(-{\bf k}),
\label{Ydecomp}
\end{eqnarray}
where $(\cdot|\cdot)$ denotes the Clebsch-Gordan coefficient
\cite{varshalo} and $\rho_i$ and $\mu_i$ are supposed to be equal $\pm
\frac12$. In the following formulae, the states with $\rho=1$ are
indexed by $++$, $--$ or $e$ depending on the quantum numbers of the
two particle direct product state. The states with zero $\rho$--spin
are indexed by $o$ respectively. The states denoted by $+-$ or $-+$ do
not have a definite $\rho$--spin. They are connected to the $e$ and
$o$ states via
\begin{equation}
{\cal Y}^{+-} =
\frac {1}{\sqrt 2}({\cal Y}^{e}
+ {\cal Y}^{o}), \qquad
{\cal Y}^{-+} =
\frac {1}{\sqrt 2}({\cal Y}^{e} - {\cal Y}^{o}).
\label{Geven}
\end{equation}

The wave function in the form of eqs.~(\ref{reldp}-\ref{Ydecomp}) is
usually referred to as direct product form. However, we find it more
convenient to work in the matrix scheme and interprete the Dirac index
of the first spinor enumerating rows and the index of the second
spinor enumerating columns.  The matrix representation of the
Bethe--Salpeter wave functions can be obtained from the direct product
form eq.(\ref{reldp}) by transposing the spinor of the second
particle. In the rest frame of the system this reads for the basis
spinors \beq {U_{\mu_1}^{(\pm)}}^{(1)}({\bf k})\;
{U_{\mu_2}^{(\pm)}}^{(2)}(-{\bf k})\; \rightarrow \;
{U_{\mu_1}^{(\pm)}}^{(1)}({\bf k})\; {U_{\mu_2}^{T(\pm)}}^{(2)}(-{\bf
  k}),
\label{replace}
\eeq
which is now a $4\times 4$ matrix in the two particle spinor space.
So, the two particle Bethe--Salpeter wave function in this basis
is represented by
\begin{eqnarray}
\Phi_{J{\cal M}}(K,k) \to
\chi_{J{\cal M}}(K,k)\;U_c=\sum \limits_{\alpha}\;
g_{\alpha} (k_0,|\bk|)\;
\Gamma_{\cal M}^{\alpha}({\bf k})\;U_C.
\label{reldmatrix}
\end{eqnarray}
where $U_C$ is the charge conjugation matrix, $U_C=i\gamma_2\gamma_0$,
and   $\Gamma_{\cal M}^{\alpha}$ is decomposed in the same way as given
for ${\cal Y}_{\cal M}^{\alpha}$ where eq.(\ref{replace}) is incorporated into
eq.(\ref{Ydecomp}). Since the matrix notation is not very familiar, we
have given an illustration of how to calculate the spin--angular
momentum part in appendix~\ref{app:spin}.

To exhibit the $\rho$-spin dependence, the spin--angular momentum
functions ${\Gamma}^\alpha_{\cal M}(\bk)$ may be replaced in general.
Using as abbreviation
\beq
\Lambda^\pm(k_1) = \frac{\hat k_1 \pm m}{\sqrt{2E_k(m+E_k)}}
\eeq
and the same formula for
$k_1\rightarrow k_2$, one may define
\beq
{\Gamma}^\alpha_{\cal M}(\bk)
\equiv {\Gamma}^{\tilde\alpha,\,\rho_1\rho_2}_{\cal M}({\bf k}),
\eeq
where
\bqn
{\Gamma}^{\tilde\alpha,\, ++}_{\cal M}(\bk) &=
&\Lambda^+(k_1)\;
\frac{1+\gamma_0}{2}\;
{\tilde \Gamma}^{\tilde\alpha}_{\cal M}(\bk)\;
\Lambda^-(k_2),
\nonumber\\
{\Gamma}^{\tilde\alpha,\, --}_{\cal M}(\bk) &=
&\Lambda^-(k_2)\;
\frac{{\lambda}(-1+\gamma_0)}{2}\;
{\tilde \Gamma}^{\tilde\alpha}_{\cal M}(\bk)\;
\Lambda^+(k_1),
\label{gf}\nonumber\\
{\Gamma}^{\tilde\alpha,\, +-}_{\cal M}(\bk) &=
&\Lambda^+(k_1)\;
\frac{1+\gamma_0}{2}\;
{\tilde \Gamma}^{\tilde\alpha}_{\cal M}(\bk)\;
\Lambda^+(k_1),
\nonumber\\
{\Gamma}^{\tilde\alpha,\, -+}_{\cal M}(\bk) &=
&\Lambda^-(k_2)\;
\frac{{\lambda}(1-\gamma_0)}{2}\;
{\tilde \Gamma}^{\tilde\alpha}_{\cal M}(\bk)\;
\Lambda^-(k_2),
\label{22}
\eqn
with $\tilde\alpha\in\{L,S,J\}$, ${\lambda}=-1$ for the $^1S_0$ channel
and ${\lambda}=+1$ for the $^3S_1-^3D_1$ channel. The expressions for
the functions ${\tilde \Gamma}^{\tilde\alpha}_{\cal M}(\bk)$ are given
in tables~\ref{tab:1s0} and \ref{tab:3s1}.

The adjoint functions are defined through
\begin{eqnarray}
{\bar {\Gamma}_{\cal M}^{\alpha}}(\bk)
=\gamma_0 \;\left[{{\Gamma}_{\cal M}^{\alpha}}
(\bk)\right]^{\dagger}\;\gamma_0,
\label{conj}
\end{eqnarray}
with the orthogonality condition given via:
\begin{eqnarray}
\int d{\hat{\bk}\
Tr \{ {{{{\Gamma}}_{\cal M}^{\alpha}}^{\dagger}(\bk)}
{\Gamma}_{{\cal M}^{\prime}}^{\alpha^{\prime}}
({\bf k})
\} = \delta_{{\cal M} {{\cal M}^{\prime}}}
\delta_{\alpha {\alpha}^{\prime}}}.
\label{ortm}
\end{eqnarray}
In addition, for identical particles the Pauli principle
holds, viz.
\begin{eqnarray}
\chi_{J{\cal M}}(K,k)=-P_{12}\chi_{J{\cal M}}(K,k)
=(-1)^{I+1}U_C\left[\chi_{J{\cal M}}(K,-k)\right]^T U_C.
\end{eqnarray}
where $I$ denotes the channel isospin.
This induces a definite  transformation property of the
radial functions $g_{\alpha} (k_0,|\bk|)$
on replacing $k_0 \rightarrow -k_0$, which is even or odd,
independend of $\alpha$. Moreover,
since the $^{3(1)}P^{\rho_1\rho_2}$ amplitudes
do not have a definite symmetry on
particle exchange, it is more convenient
to introduce $\Gamma^{P^{e}}_{\cal M}$ or $\Gamma^{P^{o}}_{\cal M}$
instead of using $\Gamma^{P^{+-}}_{\cal M}$ or $\Gamma^{P^{-+}}_{\cal M}$.
These functions do have a definite even(e) or odd(o) $\rho$-parity
which reflects a definite symmetry behaviour under particle exchange.

Subsequently, we give the explicit forms of the partial amplitude for
the $^1S_0$ and the $^3S_1$ channels.

\subsection{The $^1S_0$ channel}

For the two nucleon system in the $^1S_0$ channel the relativistic
wave function consists of four states, i.e.  $^1S_0^{++}$,
$^1S_0^{--}$, $^3P_0^{e}$, $^3P_0^{o}$, ($^3P_0^{+-}$, $^3P_0^{-+}$),
enumerated by $1,\dots,4$, respectively, in the following. The radial
functions of the partial wave decomposition of the Bethe--Salpeter
amplitude are denoted by $g_{\alpha} (k_0,|\bk|)\rightarrow\phi_i
(k_0,|{\bf k}|)$, where
\begin{equation}
k_0=\frac{(Kk)}{M},
\quad |\bk|= \sqrt{\frac {(Kk)^2}{M^2}-k^2}.
\nonumber
\end{equation}

Introducing Lorentz invariant functions
$b_i(Kk,k^2)$, the Bethe--Salpeter amplitude $\chi_{00}(K,k)$
is given by (arguments of $b_i$ suppressed)
\begin{eqnarray}
\sqrt{4\pi}\;\chi_{00}(K,k) &=&
b_1\gamma_5 +
b_2\frac {1}{m} (\hat {p}_1\gamma_5 +\gamma_5 \hat {p}_2)
\nonumber\\&&
+b_3(\frac{\hat{p}_1-m}{m}
\gamma_5 -\gamma_5 \frac {\hat {p}_2+m}{m})
+b_4 \frac{\hat{p}_1-m}{m} \gamma_5 \frac {\hat {p}_2+m}{m},
\label{covarj0}
\end{eqnarray}
where $p_{1,2}=K/2 \pm k$. The functions $b_i$ connected to $\phi_i$
are given in Appendix~\ref{app:rel}. Note,
that only $b_2$ and $\phi_4$ are odd with respect to $k_0\rightarrow
-k_0$, and all the other functions are even.

\subsection{$^3S_1- ^3D_1$ channel}
In the deuteron channel the relativistic wave function consists of
eight states, i.e. $^3S_1^{++}$, $^3S_1^{--}$,$^3D_1^{++}$,
$^3D_1^{--}$, $^3P_1^{e}$, $^3P_1^{o}$, $^1P_1^{e}$, $^1P_1^{o}$,
($^3P_1^{+-}$, $^3P_1^{-+}$, $^1P_1^{+-}$, $^1P_1^{-+}$), enumerated
by $1,\dots,8$, respectively, in the following. The radial functions
of the partial wave decomposition of the Bethe--Salpeter amplitude are
denoted by $g_{\alpha} (k_0,|\bk|)\rightarrow \psi_i (k_0,|{\bf k}|)$.

The covariant form of the Bethe Salpeter amplitude $\chi_{1{\cal
    M}}(K,k)$ with eight Lorentz invariant functions $h_i(Kk,k^2)$ is
given by (arguments of $h_i$ suppressed)
\begin{eqnarray}
\sqrt{4\pi}\;\chi_{1{\cal M}}(K,k) 
&=& 
h_1 \hat {\xi}_{\cal M} +h_2 \frac{(k \xi_{\cal M})}{m}
+h_3 \left (\frac {\hat p_1-m}{m} \hat {\xi}_{\cal M} +
\hat{\xi}_{\cal M} \frac{\hat p_2+m}{m}\right)
\label{covarj1}\\
&&+h_4 \left(\frac {\hat p_1 + \hat p_2}{m}\right)\frac {(k \xi_{\cal M})}{m}
+h_5 \left(\frac {\hat p_1-m}{m} \hat {\xi}_{\cal M} -
\hat{\xi}_{\cal M} \frac{\hat p_2+m}{m}\right)
\nonumber\\
&&
+h_6 \left(\frac {\hat p_1 - \hat p_2-2m}{m}\right)\frac {(k \xi_{\cal M})}{m}
+\frac {\hat p_1-m}{m}\left(h_7 {\hat \xi_{\cal M}}
+h_8 \frac{(k \xi_{\cal M})}{m} \right )\frac{\hat p_2+m}{m}.
\nonumber\end{eqnarray}
The relation of functions $h_i$ to $\psi_i$ are given in
appendix~\ref{app:rel}.

Note now, that $h_3$, $h_4$ and $\psi_5$, $\psi_8$ are odd, and all
the other functions are even under $k_0\rightarrow -k_0$.

\section{The electromagnetic current of the hadronic system}

\begin{figure}[h]

\unitlength=1mm
\special{em:linewidth 0.4pt}
\linethickness{0.4pt}

\hskip 30mm
\begin{picture}(50.00,70.00)
\put(20.00,60.00){\line(1,0){20.00}}
\put(40.00,60.00){\line(3,2){20.00}}
\put(20.00,60.00){\vector(1,0){10}}
\put(40.00,60.00){\vector(3,2){10}}
\put(35.00,40.00){\line(3,2){15.00}}
\put(35.00,40.00){\vector(3,2){7.5}}
\put(50.00,50.00){\line(3,-2){14.00}}
\put(50.00,50.00){\vector(3,-2){7.5}}
\put(35.00,38.00){\line(1,0){28.3}}
\put(35.00,38.00){\vector(1,0){15.00}}
\put(20.00,40.00){\line(1,0){15.00}}
\put(20.00,38.00){\line(1,0){15.00}}
\bezier{40}(25.00,41.00)(28.00,39.00)(28.00,39.00)
\bezier{40}(25.00,37.00)(28.00,39.00)(28.00,39.00)
\put(35.00,39.00){\circle*{4.00}}
\put(65.00,39.00){\circle{4.00}}
\put(66.6,40.00){\vector(1,0){8.4}}
\put(66.6,38.00){\vector(1,0){8.4}}
\put(75.00,40.00){\line(1,0){5.00}}
\put(75.00,38.00){\line(1,0){5.00}}
\bezier{40}(40.00,60.00)(42.00,60.00)(42.00,58.00)
\bezier{40}(42.00,58.00)(42.00,56.00)(44.00,56.00)
\bezier{40}(44.00,56.00)(46.00,56.00)(46.00,54.00)
\bezier{40}(46.00,54.00)(46.00,52.00)(48.00,52.00)
\bezier{40}(48.00,52.00)(50.00,52.00)(50.00,50.00)
\put(28.00,30.00){\makebox(0,0)[cc]{D: $K,{\cal M}$}}
\put(73.00,30.00){\makebox(0,0)[cc]{np: $P=K+q$}}
\put(47.00,57.00){\makebox(0,0)[cc]{$q$}}
\put(25.00,63.00){\makebox(0,0)[cc]{$k_e$}}
\put(55.00,74.00){\makebox(0,0)[cc]{$k_e^{\prime}$}}
\end{picture}

\vskip -20mm
\caption{The impulse approximation.}
\label{fig:IA}
\end{figure}
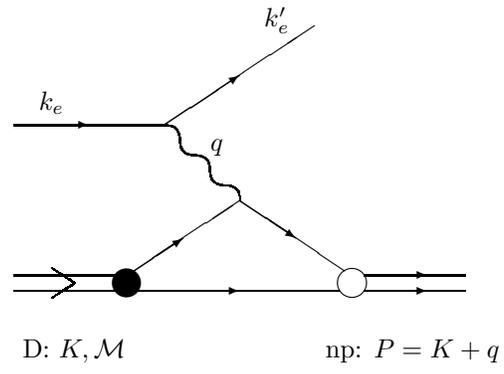

The electromagnetic current matrix element $\langle np (^1S_0)
|j_{\mu}| D {\cal M} \rangle$ will be evaluated in the relativistic
impulse approximation, see Fig.~\ref{fig:IA}.  Within the framework of
the Bethe--Salpeter (BS) approach using the Mandelstam procedure of
constructing of the electromagnetic current operator
~\cite{shebeko,lurie}, it can be written as
\begin{eqnarray}
\langle np (^1S_0) | j_{\mu} | D {\cal M} \rangle
=i \int d^4k
Tr \biggl\{ {\bar \chi_{s}}(P,p)
\Gamma^{(V)}_{\mu}(q)
\chi_{\cal M}(K,k)
(\frac{\hat K}{2}-{\hat k}+m)
\biggr\},
\label{fff}\end{eqnarray}
where $\chi_{s}(P,p)$ is the BS amplitude of the $^1S_0$ state of the
$np$-system (similar to the nonrelativistic wave function of the
continuum), and $\chi_{\cal M}(K,k)$ is the BS amplitude of the
deuteron and $p=k+q/2$. The vertex of the $\gamma NN$ interaction is
\begin{eqnarray}
\Gamma_{\mu}^{(V)}(q)=\gamma_{\mu} F_1^{(V)}(q^2)
-\frac{\gamma_{\mu} {\hat q} - {\hat q} \gamma_{\mu}}{4m}
F_2^{(V)}(q^2),
\nonumber\end{eqnarray}
which is chosen to be on the mass shell. The
isovector form factors of the nucleon appears due to summation of the
two nucleons. Instead of Dirac ($F_{1}^{(V)}(q^2)$) and Pauli
($F_{2}^{(V)}(q^2)$) form factors one may alternatively
choose Sachs form factors. The relation between them is given by
\begin{eqnarray}
 G_E(q^2) &= &F_1(q^2) + \frac{q^2}{4m^2} F_2(q^2),
\nonumber\\
 G_M(q^2) &= &F_1(q^2) + F_2(q^2).
 \label{eqn:sachs}
\end{eqnarray}

In order to extract $V(s,q^2)$ from the general expression for the
$1^+\rightarrow 0^+$ transition from eq.~(\ref{fff}), we first perform
the trace\cite{trace}.  With respect to the Lorentz structure of
 the eq.(\ref {fff}) seven integrals have to be evaluated,
which are expressed through Lorentz scalar
quantities abreviated by $(\dots)$ (the index ${\cal M}$ on the
polarization vector $\xi$ is suppressed for simplicity).
\begin{eqnarray}
{\cal I}_1&=&i{\epsilon_{\mu\alpha\beta\gamma}}\;
\xi^{\alpha}q^{\beta}K^{\gamma}
\  \int\ d^4k\ (...),
\nonumber\\
{\cal I}_2&=&i{\epsilon_{\mu\alpha\beta\gamma}}\;
\xi^{\alpha}K^{\beta}
\  \int\ d^4k\ (...)\ k^{\gamma},
\nonumber\\
{\cal I}_3&=&i{\epsilon_{\mu\alpha\beta\gamma}}\;
\xi^{\alpha}q^{\beta}
\  \int\ d^4k\ (...)\ k^{\gamma},
\nonumber\\
{\cal I}_4&=&i{\epsilon_{\mu\alpha\beta\gamma}}\;
\xi^{\delta}q^{\beta}K^{\gamma}
\  \int\ d^4k\ (...)\ k^{\alpha}k_{\delta},
\nonumber\\
{\cal I}_5&=&i{\epsilon_{\alpha\beta\gamma\delta}}\;
\xi^{\alpha}q^{\beta}K^{\gamma}
\  \int\ d^4k\ (...)\ k^{\delta}k_{\mu},
\nonumber\\
{\cal I}_6&=&i{\epsilon_{\alpha\beta\gamma\delta}}\;
\xi^{\alpha}q^{\beta}K^{\gamma}K_{\mu}
\  \int\ d^4k\ (...)\ k^{\delta},
\nonumber\\
{\cal I}_7&=&i{\epsilon_{\alpha\beta\gamma\delta}}\;
\xi^{\alpha}q^{\beta}K^{\gamma}q_{\mu}
\int\ d^4k\ (...)\ k^{\delta}.
\label{structd}
\end{eqnarray}
From inspection of eq.~(\ref {structd}) it becomes clear that besides ${\cal
  I}_1$, which needs no further consideration, two generic types of
integrals have to be evaluated, viz.
\begin{eqnarray}
&& \int\ d^4k\ (...)\ k_{\alpha}=C_1q_{\alpha}+C_2K_{\alpha},
\\
&& \int\ d^4k\ (...)\ k_{\alpha}k_{\beta}=
D_1 M^2 g_{\alpha\beta} + D_2 q_{\alpha} q_{\beta}
+ D_3 (q_{\alpha} K_{\beta} + K_{\alpha} q_{\beta})
+ D_4 K_{\alpha} K_{\beta},
\nonumber\end{eqnarray}
where we have already indicated that after integration
the expressions should depend on the external 4-momenta
only, i.e. the transition momentum $q$ and the total deuteron
momentum $K$. Due to the antisymmetric tensor, the number of
terms is reduced substantially, and all the terms with $D_i$, $i>1$
vanish. So, besides ${\cal I}_1$, it is necessary to evaluate
the following integrals:
\begin{eqnarray}
{\cal I}_2&=&i{\epsilon_{\mu\alpha\beta\gamma}}\;
\xi^{\alpha}K^{\beta} q^{\gamma}
\  \int\ d^4k\ (...)\ c_1(k,q,K),
\nonumber\\
{\cal I}_3&=&i{\epsilon_{\mu\alpha\beta\gamma}}\;
\xi^{\alpha}q^{\beta} K^{\gamma}
\  \int\ d^4k\ (...)\ c_2(k,q,K),
\nonumber\\
{\cal I}_4&=&i{\epsilon_{\mu\alpha\beta\gamma}}\;
\xi^{\alpha}q^{\beta}K^{\gamma}
\ M^2 \int\ d^4k\ (...)\ d(k,q,K),
\nonumber\\
{\cal I}_5&=&i{\epsilon_{\alpha\beta\gamma\mu}}\;
\xi^{\alpha}q^{\beta}K^{\gamma}
\ M^2 \int\ d^4k\ (...)\ d(k,q,K),
\end{eqnarray}
and ${\cal I}_6={\cal I}_7=0$; $c_i$ and $d$ are scalar functions
given in Appendix~\ref{app:V}.

Integration and comparison with the structure of the transition matrix
element given in  eq.(\ref{formc}) finally leads to determine
$V(s,q^2)$. The result is
\begin{equation}
V(s,q^2) = V_1(s,q^2)F_1^{(V)}(q^2)
+ V_2(s,q^2)F_2^{(V)}(q^2).
\label{eqn:V}
\end{equation}
The expressions for $V_{1,2}(s,q^2)$ are rather lengthy and therefore
shifted to appendix~\ref{app:V}.

\section{Relativistic plane wave impulse approximation}

It is instructive to approximate the final $np$ state by a plane wave
in order to get a better impression of the terms that contribute to
the form factor $V(s,q^2)$ defined in eq.~(\ref{formc}) and evaluated
in the last section.  We use the covariant form of the BS amplitude of
the $np$-system and the deuteron. In the plane wave approximation the
$^1S_0$ channel radial amplitudes in the rest frame of the pair are
given by
\begin{eqnarray}
\phi_{\Sp}(p_0,|\bp|) &=
&\frac{1}{\sqrt{4\pi}}\;\frac{1}{\bp^{*\, 2}}\;\delta(p_0)\;
\delta(|\bp|-{|\bp^{*}|}),
\nonumber\\
\quad \phi_{\Sm}(p_0,|\bp|) &= &0,
\nonumber\\
\quad \phi_{\PAe}(p_0,|\bp|) &= &0,
\nonumber\\
\quad \phi_{\PAo}(p_0,|\bp|) &= &0,
\label{31}\end{eqnarray}
where $p_0,|\bp|$  stand for the relative momentum of the pair in the
rest frame and $|\bp^{*}|=\sqrt{s/4-m^2}$.

The calculations have been  done in the laboratory system. Writing the
$\delta$--functions in eq.~(\ref{31}) as
\begin{eqnarray}
&&\delta(\frac{Pp}{\sqrt{s}})=\frac{\sqrt{s}}{M+q_0}\delta(k_0-\frac
{\bk\bq-1/2Mq_0-1/2q^2}{M+q_0}),
\nonumber\\
&&
\delta(\sqrt{s/4-m^2}-\sqrt{-(k+q/2)^2})=\;\frac{\sqrt{s/4-m^2}(M+q_0)}
{E_k|\bk||\bq|}\;\delta(\cos{\theta_{\bk}}-\frac{s-2E_k(M+q_0)}{2|\bk||\bq|}),
\label{32}\end{eqnarray}
one can perform integration in eq.(\ref{eqn:V}) and set
\begin{eqnarray}
k_0=k_0^{\star}=\frac{M}{2}-E_k,
\quad\quad
\cos{\theta_{\bk}}=\frac{s-2E_k(M+q_0)}{2|\bk||\bq|}.
\label{33}
\end{eqnarray}

The condition $|\cos{\theta_{\bk}}| \leq 1$ restricts the integration
limits of $|\bk|$ and the following expression for $V_{1,2}(s,q^2)$
can be obtained
\begin{eqnarray}
V_{1,2}(s,q^2) &=& \frac{1}{\sqrt{4\pi}}\;
\frac{\sqrt{s}}{2|\bq||\bp^{*}|}\;
\int\limits_{|{\bk}|_{min}}^{|{\bk}|_{max}} d|{\bk}|\;
\frac{|{\bk}|}{E_k}\;I_{1,2}(q^2,s,|{\bk|}),
\label{ff}
\end{eqnarray}
where
\begin{eqnarray}
|{\bk}|_{max,min}=\Biggl|\sqrt{1+\frac{\bq^2}{s}}\;
\sqrt{\frac{s}{4}-m^2}\;\pm\;
\frac{|\bq|}{2}\Biggr|,
\quad\quad
|\bq|=\frac{\sqrt{((M^2+s-q^2)^2-4sM^2)}}{2M}
\end{eqnarray}
and

\begin{eqnarray}
I_{1}(s,q^2,|{\bk}|) &=&
-\frac{a}{(m+E_k)}
\bigl[s(s+4E_k^2)-4m{\bq}^2(m+2E_k)-4sE_kb\bigr]
\frac{g_{S}(k_0^{\star},|{\bk}|)}{(M-2E_k)}
\nonumber\\
&&-\frac{\sqrt{2}a}{2{\bk}^2}\bigl[s(m+2E_k)
(s+4E_k^2-4E_kb)+4m{\bq}^2(E_k(E_k+2m)+{\bk}^2)
\bigr]
\frac{g_{D}(k_0^{\star},|{\bk}|)}{(M-2E_k)}
\nonumber\\
&&-\frac{\sqrt{3}a}{2M|{\bk}|}\bigl[s(s-4E_k^2)-4m^2{\bq}^2
\bigr]
(g_{t}^{e}(k_0^{\star},|{\bk}|)-g_{t}^{o}(k_0^{\star},|{\bk}|)),
\label{43}\\
I_{2}(s,q^2,|{\bk}|) &=&
\frac{a}{2m(m+E_k)}
\bigl[s[s(M-4(E_k+m))-4E_kM(E_k+2m)]+4m^2M{\bq}^2
\nonumber\\
&&+4s(Mm+2E_k(E_k+m))b\bigr]
\frac{g_{S}(k_0^{\star},|{\bk}|)}{(M-2E_k)}
\nonumber\\
&&-\frac{\sqrt{2}a}{4m{\bk}^2}\bigl[s^2(M(E_k+2m)-4{\bk}^2)
+4MsE_k(2m(E_k+m)-E_k^2)
\nonumber\\
&&-4s(Mm(m+2E_k)-2E_k{\bk}^2)b+4Mm^2{\bq}^2(E_k+2m)
\bigr]
\frac{g_{D}(k_0^{\star},|{\bk}|)}{(M-2E_k)}
\nonumber\\
&&-\frac{\sqrt{3}a}{M|{\bk}|}(s-2bE_k)(s-Mb)
(g_{t}^{e}(k_0^{\star},|{\bk}|)-g_{t}^{o}(k_0^{\star},|{\bk}|))
\nonumber\\
&&-\frac{\sqrt{3}\sqrt{2}a}{4m|{\bk}|}\bigl[s(s+4E_k^2-
4bE_k)+4m^2{\bq}^2\bigr]
(g_{s}^{e}(k_0^{\star},|{\bk}|)-g_{s}^{o}(k_0^{\star},|{\bk}|))
\label{44}
\end{eqnarray}
with $a=1/4Ms{\bq}^2$, $b=\sqrt{s+{\bq}^2}$, $k_0^{\star}=M/2-E_k$,
and $g_{\alpha}(k_0,|\bk|)$ are the vertex functions of the partial
states of the deuteron. The indices $S,D$ denote the $^3S_{1}^{+}$,
$^3D_{1}^{+}$ states, resp., and index $s(t)$ denotes the singlet
(triplet) states and $o(e)$ the odd (even) state with respect to the
$\rho$-spin.

As it can be seen from eqs.~(\ref{43}-\ref{44}) only the $(++)$ or
$(-+)$ components of the deuteron function are present. This can
easily be understood by writing the matrix element eq.~(\ref{fff}) in
the laboratory system and using wave functions of the form
eq.~(\ref{22}). The expression for the inverse propogator in
eq.~(\ref{fff}) is of the following form:
\begin{eqnarray}
\frac{\hat K}{2}-{\hat k}+m = \frac{1}{2E_k}
\bigl[({\hat k}_1-m)(\frac{M}{2}-k_0+E_k)
+({\hat k}_2+m)(\frac{M}{2}-k_0-E_k)
\bigr],
\label{prop}\end{eqnarray}
with $k_1=(E_k,\bk)$ and $k_2=(E_k-\bk)$.
Thus, for the $(--)$ component we get
\begin{eqnarray}
&& \chi_{\cal M}(K,k)(\frac{\hat K}{2}-{\hat k}+m) \sim
\nonumber\\
&& ({\hat k}_2-m) \frac{1-\gamma_0}{2} {\tilde \Gamma}_{\cal M}(\bk)
({\hat k}_1+m)
\bigl[({\hat k}_1-m)(\frac{M}{2}-k_0+E_k)
+({\hat k}_2+m)(\frac{M}{2}-k_0-E_k)
\bigr] = 0
\label{zero}\end{eqnarray}
because of $k_0=M/2-E_k$ and $({\hat k}_1+m)({\hat k}_1-m)=0$.  The
same is valid for the $(+-)$ components.

Then the transition form factor $V(s,q^2)$ has a specific simple form
in the case of $s=4m^2$, viz.
\begin{eqnarray}
V(4m^2,q^2) 
&=& \frac{1}{\sqrt{4\pi}}\;
\Biggl\{
\frac{1}{2Mm}\;
\bigl(F_{1}^{(V)}(q^2)+F_{2}^{(V)}(q^2)\bigr)\;
\frac{ g_S(k_0^{\star},|{\bq}|/2)-
\frac{1}{\sqrt{2}}g_D(k_0^{\star},|{\bq}|/2)}
{M-2\sqrt{m^2+{\bq}^2/4}}
\label{eqn:sfourm}
\\
&+& \frac{\sqrt{3}}{2M^2|{\bq}|}
\Bigl(F_{1}^{(V)}(q^2)+\biggl[1-\frac{M}{2m}\sqrt{1+\frac{\bq^2}{4m^2}}
\biggr]F_{2}^{(V)}(q^2)\Bigr)
\bigl(g^{e}_t(k_0^{\star},|{\bq}|/2)
-g^{o}_t(k_0^{\star},|{\bq}|/2)\bigr)
\Biggr\},
\nonumber\end{eqnarray}
where $k_0^{\star}=M/2-\sqrt{m^2+\bq^2/4}$.

\section{Non--relativistic interpretation and the pair current}

Disintegration of the deuteron at threshold has been a cornerstone to
establish subnucleonic degrees of freedom in the electromagnetic
current operator. Compared to the data the non--relativistic impulse
approximation fails completely, and additional corrections, viz. two
body currents have been introduced.  On the other hand, relativistic
impulse approximation partially includes two body currents. Recently,
this has been shown explicitly in the context of the light front
approach to the two nucleon system~\cite{karmanov}. Here we consider
the covariant approach and like to offer a motivation why the two body
currents are related to the $P$--wave components which are not present
in the nonrelativistic treatment.

As a demonstration we compare the result of the relativistic
calculation of the deuteron disintegration amplitude for the case
$p^*=0$ with the nonrelativistic one. In this case the nonrelativistic
formula is usually parameterized in term of the following reduced
matrix element~\cite{mathiot,foldy}
\begin{eqnarray}
\langle ^1S_0 || T^{Mag} || D \rangle 
&=&
i\frac{|\bq|}{2m}\frac{1}{\sqrt{2\pi}}
\Biggl\{
\bigl(F_{1}^{(V)}(q^2)+F_{2}^{(V)}(q^2)\bigr)
\int_0^{\infty} |\br| d|\br|
\Bigl[ u_S(|\br|) j_0\Bigl(\frac{|\br||\bq|}{2}\Bigr)-
\frac{1}{\sqrt{2}} u_D(|\br|) j_2\Bigl(\frac{|\br||\bq|}{2}\Bigr) \Bigr]
\nonumber\\
&+& \xi^{(V)}(q^2) \frac{g^2_{\pi NN}}{4\pi}\frac{2}{m|\bq|}
\int_{0}^{+\infty} d|\br| \frac{e^{-\mu|\br|}}{|\br|} (1+\mu|\br|)
\left(u_{S}(|\br|)
+\frac{1}{\sqrt{2}}u_{D}(|\br|)\right)
j_1\Bigl(\frac{|\br||\bq|}{2}\Bigr)  \Biggr\}.
\label{eqn:NR}\end{eqnarray}
where $u_S$ and $u_D$ are the nonrelativistic $S-$ and $D-$wave
functions. The first term in eq.~(\ref{eqn:NR}) corresponds to the
nonrelativistic impulse approximation and the second term -- to
mesonic exchange currents. Within the nonrelativistic calculations the
nucleon form factor $\xi^{(V)}(q^2)$ in eq.~(\ref{eqn:NR}) has been
subjected to discussion. Two possibilities have been considered, i.e.
$\xi^{(V)}(q^2)=F_1^{(V)}(q^2)$~\cite{f1ge} and
$\xi^{(V)}(q^2)=G_E^{(V)}(q^2)$~\cite{Geschool}.  Within the covariant
approach used here, form factor $\xi^{(V)}(q^2) $ cannot be chosen
freely, that will be explained below.

In the limit $k^{\star}_0\rightarrow 0$ it is possible to relate the
nonrelativistic wave functions to the relativistic vertex function,
viz.~\cite{bondarenko}
\begin{equation}
g_{S,D}(k_0^* \to 0,|\bk|) \to \alpha\;m\;(2E_k-M)\;u_{S,D}(|\bk|),
\label{eqn:link}
\end{equation}
with $\alpha=4\pi\sqrt{2M}/m$.  Inserting eq.~(\ref{eqn:link}) into
eq.~(\ref{eqn:sfourm}) one recovers the structure of the
nonrelativistic result as given in eq.~(\ref{eqn:NR}).  The term
proportional to $(F_1+F_2)$ is determined by the positive energy
components of the Bethe--Salpeter amplitudes, $g_S$ and $g_D$. We
argue that they may be interpreted as the nonrelativistic impulse
approximation. Then the mesonic exchange current contribution
eq.(\ref{eqn:NR}) may be related to the remaining terms that depend on
the $P$--wave components only.

Since within the covariant approach the sets of form factors \{$F_1$,
$F_2$\} and \{$G_E$, $G_M$\} are completely equivalent, the discussion
whether to use $F_1$ or $G_E$ in the form factors of the mesonic
exchange part, is obsolete. Our result is
\begin{eqnarray}
\xi^{(V)}(q^2)&=&
\left(F_{1}^{(V)}(q^2)+\biggl[1-\frac{M}{2m}\sqrt{1+\frac{\bq^2}{4m^2}}
\biggr]F_{2}^{(V)}(q^2)\right)\nonumber\\& =&
\left(G^{(V)}_E(q^2)+[1-\frac{M}{2m}\sqrt{1+\frac{\bq^2}{4m^2}}-
\frac{q^2}{4m^2}]F_2^{(V)}(q^2) \right).
\end{eqnarray}
This is not the case for nonrelativistic approximations where
covariance is broken and higher order corrections in $q/m$ are not
treated properly. In this case some physics aspects may be described
better by using either one of them~\cite{f1ge,Geschool}.  With the
definition of eq.~(\ref{eqn:sachs}) and using $G_E$ we may express the
form factor dependent part of the second term in
eq.~(\ref{eqn:sfourm}) belonging to the mesonic exchange currents. For
simplicity we neglect the deuteron binding energy (viz.  $M=2m$), and
$O(q^4)$ terms in the expansion of the square root and write
\begin{eqnarray}
\xi^{(V)}(q^2)
&=& F_{1}^{(V)}(q^2)+ \frac{q^2}{8m^2}F_{2}^{(V)}(q^2)+ O(q^4)
\nonumber\\
&=& G_{E}^{(V)} - \frac{q^2}{8m^2}F_{2}^{(V)}(q^2)+O(q^4).
\label{eqn:pairsachs}
\end{eqnarray}

\section{ Conclusion and Outlook}

Although the nonrelativistic limit of the Bethe--Salpeter equation
does not exist in the general case, it is possible to connect
relativistic and nonrelativistic calculations of the deuteron
disintegration amplitude.  The nonrelativistic impulse approximation
corresponds to the states with positive energies.  Corrections due to
the $P$--components are presumably connected to mesonic exchange
currents. In further communications we shall give an analytic proof
that the main contribution of the $P$--states to the amplitude
coincides with the non relativistic pair current.  This conclusion is
supported by the calculation of the magnetic moment, where it has been
shown that the size and sign of the $P$--state contribution are the
same as in the pair current contribution within the nonrelativistic
framework.

It has been shown that the Bethe--Salpeter approach allows to
calculate the amplitude of the disintegration process in a covariant
form.  This is a great advantage, since it gives an opportunity to
consider some old problems from a new point of view.  One of the
problems is gauge independence of the current matrix element. It has
not been solved for the non relativistic treatment. Therefore the
problem of choosing the correct form factor in meson exchange currents
stays persistant. A fully relativistic approach is free from these
biases. Another interesting result unseen in the non relativistic
approach is that simple relations between polarization observables can
be achieved.

\begin{figure}[h]
\epsfxsize=0.8\textwidth
\epsfbox{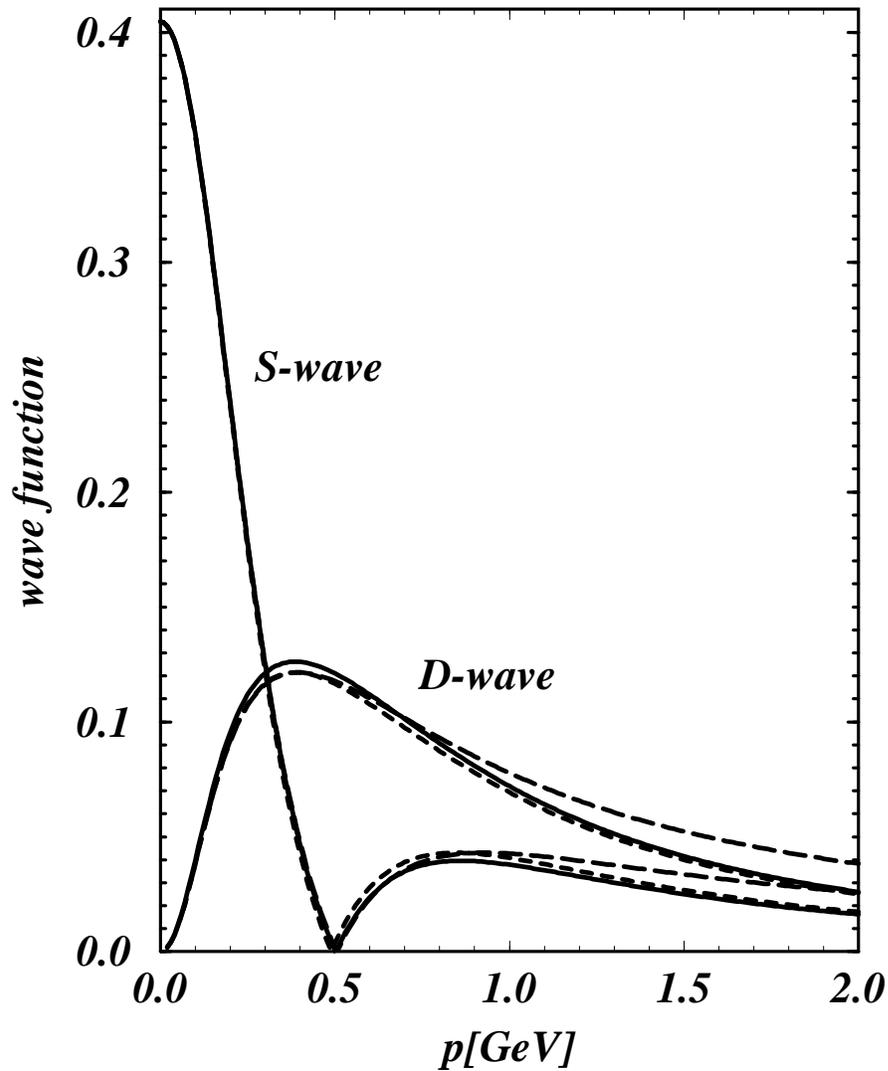}
\caption{The $k_0$ dependence of the vertex functions. The short dashed
  line is the relativistic vertex function with $k_0=0$, the long
  dashed line is the relativistic vertex function with $k_0=M/2-E_k$,
  the solid line is the nonrelativistic vertex function. The functions
  are fron the Graz II potential [24]}
\label{fig:1}
\end{figure}

Usually the Bethe--Sapleter equation is solved after Wick rotation.
The formula of the present calculation has shown that the dependence
of the vertex function on the variable $k_0$ is sufficient. This is
similar to calculations in the context of the $pd$ elastic
scattering~\cite{tjon:pd} where it has been motivated that the
dependence on $k_0$ does not substantially influence the results by
choosing $k_0=0$.  However, in the case of deuteron disintegration in
the relevant regime of momentum transfer, $q^2\le 3$GeV$^2$, the
values of $k_0$ vary from 0 to 1 GeV. The dependence of the vertex
function on $k_0$ is displayed in Figure~\ref{fig:1}. The difference
of the vertex functions with $k_0=0$ and $k_0=k^{\star}\neq 0$ can be
as large as 20\%.

\section{Acknowledgment}  

The authors gratefully acknowledge discussions with V.A. Karmanov and
L.P. Kaptari.  This work is supported by a collaborative grant of
Deutsche Forschungsgemeinschaft and Russian Foundation of Fundamental
Research. The possibilities to communicate between the Far Eastern
State University and Rostock University have been strongly improved by
a NATO infrastructure grant.

\newpage
\appendix

\section{Calulation of spin--orbit dependence}\label{app:spin}

Here is an example
how to calculate the spin -- angular momentum
part of the wave function for the $^3 S_1^{++}$ -
state where the spectroscopic notation $^{2S+1}L^{\rho}
_{J}$ of ref.~\cite{kubis} is used. In this case
\begin{eqnarray}
\sqrt{4 \pi}\; {\Gamma}^{^3 S_1^{++}}_{\cal M}( \mbf{k})U_C
& = &\sum \limits _{\mu_1 \mu_2}\;
(\Hh \mu_1 \Hh \mu_2 |1{\cal M})\; u_{\mu_1}( \mbf{k})\;
u_{\mu_2}^T(-\mbf{k})\nonumber\\
&=&{{\hat{\cal L}}}( \mbf{k})\sum \limits _{\mu_1 \mu_2}
\;(\Hh \mu_1 \Hh \mu_2 |1{\cal M})
\;\Bigl(\begin{array}{c} \chi_{\mu_1} \\ 0 \end{array} \Bigr)
\;(\begin{array}{cc} \chi_{\mu_2}^T &0 \end{array} )
\;{{\hat{\cal L}}}^T(-\mbf{k})\nonumber\\
&=&{{\hat{\cal L}}}( \mbf{k})
\;\Biggl( \begin{array}{cc}
\sum \limits_{\mu_1 \mu_2}
(\Hh \mu_1 \Hh \mu_2 |1{\cal M})
\chi_{\mu_1} \chi^T_{\mu_2} &0 \\ 0&0 \end{array}\Biggr)
\;{{\hat{\cal L}}}^T(-\mbf{k})\nonumber\\
&=&{{\hat{\cal L}}}( \mbf{k})
\;\frac {1+\gamma_0}{2} \frac 1{\sqrt2}
\;\Biggl( \begin{array}{cc} 0 & -\mbf{\sigma}\cdot \mbf{\xi}_{\cal M}\\
 \mbf{\sigma}\cdot\mbf{\xi}_{\cal M} &0 \end{array}\Biggr)
\;\Biggl( \begin{array}{cc} 0 & -i \sigma_2\\
-i\sigma_2 &0 \end{array}\Biggr)
\;{{\hat{\cal L}}}^T(-\mbf{k})\nonumber\\
&=&{{\hat{\cal L}}}( \mbf{k})
\;\frac {1+\gamma_0}{2} \frac 1{\sqrt2}
\;(-\mbf{\gamma}\mbf{\xi}_{\cal M})\;{{\hat{\cal L}}}( \mbf{k})
U_C\nonumber\\
& =&\frac{1}{2E_k(m+E_k)}\frac{1}{\sqrt{2}}\;(m+{\hat k_1})\;
\frac{1+\gamma_0}{2}\; \xi_{\cal M}\;(m-{\hat k_2})\;U_C,
\nonumber
\end{eqnarray}
Here we have used the identity $\sqrt{2}\sum \limits_{\mu_1 \mu_2}
(\Hh \mu_1 \Hh \mu_2 |1{\cal M}) \chi_{\mu_1} \chi^T_{\mu_2}=
(\mbf{\sigma}\cdot\mbf{\xi}_{\cal M}) \;({i \sigma_2})$, where
$k_2=(E_k,-\bk)$, $\mbf{\xi}_{\cal M}$ is the polarization vector of
the spin-1 composite system with the components in the rest frame
given by
\begin{eqnarray}
\mbf{\xi}_{+1}=(-1,-i,0)/\sqrt{2}, \quad
\mbf{\xi}_{-1}=(1,-i,0)/\sqrt{2}, \quad
\mbf{\xi}_{0}=(0,0,1).
\label{vecpol}
\end{eqnarray}
and the 4-vector $\xi_{\cal M} = (0,\mbf{\xi}_{\cal M})$.

\section{Relation of covariant form to partial wave expansion}
\label{app:rel}
The relation between functions $h_i$ of the covariant expansion to
the functions $\phi_i$ of the partial wave expansion is given by,
for the $^1S_0$ channel:
\begin{eqnarray}
&&b_1=
-\frac{\sqrt{2}}{16}a_1 [D_1^{-}\phi_1-D_1^{+}\phi_2]
-\frac{1}{8}a_1a_2D_2\phi_3+\frac{1}{M}a_2k_0\phi_4\\
&&b_2=-\frac{1}{4}a_2\phi_4\\
&&b_3=
\frac{\sqrt{2}}{2}a_1m^2[\phi_1-\phi_2]
-\frac{1}{2}a_1a_2(E_{k}^2-2m^2)\phi_3
+\frac{1}{2M}a_2k_0\phi_4
\\
&&b_4=
-\frac{\sqrt{2}}{4}a_1m^2[\phi_1-\phi_2]
-a_1a_2m^2\phi_3
\label{eqn:H}
\end{eqnarray}
where
\begin{eqnarray*}
a_1&=&1/ME_{k},\\
a_2&=&m/|\bk|,\\
D_1^{\pm}&=&(M\pm 2E_{k})^2-4(4m^2+k_0^2),\\
D_2&=&M^2+12E^2_{k}-16m^2-4k_0^2,
\end{eqnarray*}
for the $^3S_1-^3D_1$ channel:
\begin{eqnarray}
&&h_1=
\frac{\sqrt{2}}{16}a_1 [D_1^{-}\psi_1+D_1^{+}\psi_2]
-\frac{1}{16}a_1 [D_1^{-}\psi_3+D_1^{+}\psi_4]
+\frac{\sqrt{6}}{2M}a_2 k_0\psi_5
-\frac{\sqrt{6}}{16}a_1a_2D_2\psi_6
\\
&&h_2=
\frac{\sqrt{2}}{16}a_1a_3 [D_3^{-}\psi_1+D_3^{+}\psi_2]
+\frac{1}{16m}a_1a_2^2 [D_4^{+}\psi_3+D_4^{-}\psi_4]
-\frac{\sqrt{6}}{2M}a_2 k_0\psi_5
\\
&&-\frac{\sqrt{6}}{16}a_1a_2D_5\psi_6+
\sqrt{3}\frac{m}{E_{k}}a_2\psi_7
\\
&&h_3=-\frac{\sqrt{6}}{8}a_2\psi_5
\\
&&h_4=
\frac{\sqrt{2}}{4}a_1a_3mk_0[\psi_1-\psi_2]
-\frac{1}{4}a_1a_2^2(E_{k}+2m)k_0[\psi_3-\psi_4]
-\frac{\sqrt{3}}{2}a_1a_2mk_0\psi_7
+\frac{\sqrt{3}}{2}\frac{m}{M}a_2\psi_8
\\
&&h_5=
-\frac{\sqrt{2}}{2}a_1m^2[\psi_1+\psi_2]
+\frac{1}{2}a_1m^2[\psi_3+\psi_4]
+\frac{\sqrt{6}}{4M}a_2k_0\psi_5
-\frac{\sqrt{6}}{4}a_1a_2(E_{k}^2-2m^2)\psi_6
\\
&&h_6=\frac{\sqrt{2}}{8}a_1a_3m [D_6^{-}\psi_1+D_6^{+}\psi_2]
+\frac{1}{8}a_1a_2^2 [D_7^{+}\psi_3+D_7^{-}\psi_4]
-\frac{\sqrt{6}}{2}a_1a_2m^2 \psi_6
+\frac{\sqrt{3}}{4}\frac{m}{E_{k}}a_2\psi_7
\\
&&h_7=
\frac{\sqrt{2}}{4}a_1m^2 [\psi_1+\psi_2]
-\frac{1}{4}a_1m^2 [\psi_3+\psi_4]
-\frac{\sqrt{6}}{4}a_1a_2m^2 \psi_6
\\
&&h_8=
-\frac{\sqrt{2}}{4}a_1a_3m^2 [\psi_1+\psi_2]
-\frac{1}{4}a_1a_2^2 m(2E_{k}+m)[\psi_3+\psi_4]
+\frac{\sqrt{6}}{4}a_1a_2m^2 \psi_6
\end{eqnarray}
where
\begin{eqnarray*}
a_3&=&m/(E_{k}+m),\\
D_3^{\pm}&=&(M \pm 2E_{k} \pm 4m)^2-4k_0^2,\\
D_4^{\pm}&=&[(M\pm 2E_{k})^2-4k_0^2](2E_{k}+m)-16(m\pm M)\bk^2,\\
D_5&=&M^2-4E_{k}^2+16m^2-4k_0^2,\\
D_6^{\pm}&=&4m+2E_{k}\pm M,\\
D_7^{\pm}&=&\pm M(E_{k}+2m)+2(2m^2+2mE_{k}-E_{k}^2).
\end{eqnarray*}

\section{The transition form factor}
\label{app:V}

The eq.~(\ref{eqn:V}) for $V(s,q^2)$ given in section III is
$$
V(s,q^2) = V_1(s,q^2)F_1^{(V)}(q^2)
+ V_2(s,q^2)F_2^{(V)}(q^2).
$$
The expressions for $V_{1,2}(s,q^2)$ are
\begin{eqnarray}
V_1(s,q^2) &=& \int d^4k
\left(V_1^{(1)}(s,q^2,k)+V_2^{(1)}(s,q^2,k)
+V_3^{(1)}(s,q^2,k)+V_4^{(1)}(s,q^2,k)\right)\\
V_2(s,q^2) &= &\int d^4k
\left(V_1^{(2)}(s,q^2,k)+V_2^{(2)}(s,q^2,k)
+V_3^{(2)}(s,q^2,k)+V_4^{(2)}(s,q^2,k)\right)
\end{eqnarray}

where
\begin{eqnarray}
V_1^{(1)} =&& 4\omega_1 c_1b_1
\Bigl[
h_3+h_5+2h_7
\Bigr]
\nonumber\\
V_2^{(1)} =&&\omega_1 b_2
\Bigl[
-2\Bigl(2(c_1-c_2)+1\Bigr)h_1 + 8c_2h_3 - 4\omega_2 dh_4
+8\Bigl(2c_1-c_2+1\Bigr)h_5
-4\omega_2 dh_6
\nonumber\\
&& + \frac{\omega_3}{2}
\Bigl(2(4k^2+12m^2+M^2-4(Kk))c_1+2(4k^2-4m^2-4(Kk)+M^2)c_2
\nonumber\\
&& + 4k^2+12m^2+M^2-4(Kk)\Bigr)h_7 - 8\omega_2 dh_8
\Bigr]
\nonumber\\
V_3^{(1)} =&&\omega_1 b_3
\Bigl[
-2\Bigl(2(c_1-c_2)+1\Bigr)h_1 -8\Bigl(2c_1-c_2\Bigr)h_3
-4\omega_2 dh_4 +8\Bigl(1-c_2\Bigr)h_5
- 4\omega_2 dh_6
\nonumber\\
&& - \frac{\omega_3}{2} \Bigl(
2(M^2+12m^2+4k^2-4(Kk))c_1+2(4m^2-4k^2-M^2+4(Kk))c_2
\nonumber\\
&& - M^2-12m^2-4k^2+4(Kk)\Bigr)h_7 - 8\omega_2 dh_8
\Bigr]
\nonumber\\
V_4^{(1)} =&&\omega_1 b_4
\Bigl[
-4\Bigl(2(c_1-c_2)+1\Bigr)h_1-
\omega_3 \Bigl((12m^2+M^2-8(kq)+4k^2-4(Kk)+4(Kq))c_1
\nonumber\\
&& + 2(M^2-4m^2-4k^2)c_2-4(Kk)+8k^2-16dM^2)\Bigr)h_3
-8\omega_2 dh_4
\nonumber\\
&& + \omega_3
\Bigl((12m^2-4(Kq)+M^2+4k^2+8(kq)-4(Kk))c_1+
4(2(Kk)-4m^2-M^2)c_2
\nonumber\\
&& + M^2+12m^2-4k^2+16dM^2\Bigr)h_5
-8\omega_2 dh_6
+4\Bigl(2(kq)-(Kq)\Bigr)c_1
\nonumber\\
&& + 2\omega_3 \Bigl(
(4(Kk)-4m^2+4k^2-3M^2)c_2-4k^2+M^2+4m^2+16dM^2\Bigr)h_7
\nonumber\\
&& - \omega_2 \Bigl(M^2+12m^2+4k^2-4(Kk)\Bigr)dh_8
\Bigr]
\nonumber
\end{eqnarray}

\begin{eqnarray}
V_1^{(2)}=&&\omega_1 b_1
\Bigl[
-\Bigl(1-2c_2\Bigr)h_1+4c_2h_3-2\omega_2 dh_4+4\Bigl(1-c_2\Bigr)h_5
-2\omega_2 dh_6
\nonumber\\
&& +\frac{\omega_3}{4}\Bigl(2(4k^2-4m^2-4(Kk)+M^2)c_2
+4k^2+12m^2+M^2-4(Kk)\Bigr)h_7
-4\omega_2 dh_8
\Bigr]
\nonumber\\
V_2^{(2)}=&&\omega_1 b_2
\Bigl[
4c_2h_1-2\omega_2 dh_2
+2\omega_3 \Bigl((4(kq)+q^2)c_1+4c_2k^2-2k^2+6dM^2\Bigr)h_3
-4\omega_2 dh_4
\nonumber\\
&& + 2\omega_3
\Bigl((4(kq)+q^2)c_1+(4(Kk)-M^2-4m^2)c_2-2k^2+6dM^2\Bigr)h_5
+4\omega_2 dh_6
\nonumber\\
&& +\omega_3 \Bigl(16c_1(kq)+4c_1q^2+(12(Kk)-4m^2+4k^2-3M^2)c_2
-8k^2+24dM^2\Bigr)h_7
\nonumber\\
&& -\frac{\omega_2}{2}\Bigl(M^2-4m^2+4k^2-4(Kk)\Bigr)dh_8
\Bigr]
\nonumber\\
V_3^{(2)}=&&\omega_1 b_3
\Bigl[
4\Bigl(1-c_2\Bigr)h_1-2\omega_2 dh_2
+2\omega_3 \Bigl((2(Kq)+q^2)c_1+(M^2-4m^2)c_2
\nonumber\\
&& -2(Kk)+2k^2-2dM^2\Bigr)h_3
+4\omega_2 dh_4
+\omega_3 \Bigl(
2(2(Kq)+q^2)c_1+2(4m^2+M^2)c_2
\nonumber\\
&&-12m^2-M^2-4dM^2\Bigr)h_5+12\omega_2 dh_6+
\Bigl(4(q^2+2(Kq))c_1
\nonumber\\
&& +(3M^2+4m^2-4k^2+4(Kk))c_2
-8m^2-2M^2-8dM^2\Bigr)h_7
\nonumber\\
&&+\frac{\omega_2}{2}\Bigl(28m^2+M^2+4k^2-4(Kk)\Bigr)dh_8
\Bigr]
\nonumber\\
V_4^{(2)}=&&\omega_1 b_4
\Bigl[
\frac{\omega_3}{4}
\Bigl(8(2(kq)-(Kq))c_1+2(4k^2-4m^2+4(Kk)-3M^2)c_2
\nonumber\\
&& -12k^2+M^2+12m^2+4(Kk) + 32dM^2\Bigr)h_1
-4\omega_2 dh_2 +
\Bigl(4(q^2+(Kq)+2(kq))c_1
\nonumber\\
&&+(M^2+4k^2-4m^2+4(Kk))c_2 - 4(Kk)+8dM^2\Bigr)h_3
\nonumber\\
&& -\frac{\omega_2}{2}\Bigl(M^2-4m^2+4k^2-4(Kk)\Bigr)dh_4
-\omega_3 \Bigl(4(2(kq)-3(Kq)+4(kq)-q^2)c_1
\nonumber\\
&&+(4k^2+4(Kk)-4m^2-7M^2)c_2
-8k^2+2M^2+8m^2+4(Kk)+24dM^2\Bigr)h_5
\nonumber\\
&&+\frac{\omega_2}{2}\Bigl(28m^2+M^2+4k^2-4(Kk)\Bigr)dh_6
+\frac{\omega_4}{2}
\Bigl(
(M^2(Kq)+12q^2m^2+2M^2(kq)
\nonumber\\
&& -8(kq)(Kk)+q^2M^2-8(kq)m^2+28(Kq)m^2+4q^2k^2+4(Kq)k^2+8k^2(kq)
\nonumber\\
&& -4(Kq)(Kk)-4q^2(Kk))c_1
+2(16m^4+16k^4+M^4-16(Kk)^2-32k^2m^2
\nonumber\\
&& +8M^2k^2+56M^2m^2)c_2
+96k^2m^2-8M^2k^2
-64(Kk)m^2-40M^2m^2-M^4
\nonumber\\
&& +16(Kk)^2-16k^4-80m^4
+16(4k^2+M^2-4(Kk)-20m^2)M^2d\Bigr)h_7
\nonumber\\
&& +\omega_5\Bigl(M^2-4(Kk)+4k^2+12m^2\Bigr)dh_8
\Bigr]
\nonumber
\end{eqnarray}
with the abreviations
$$
\omega_1=1/m,\quad \omega_2=M^2/m^2,\quad \omega_3=1/m^2,\quad
\omega_4=1/m^4,\quad \omega_5=M^2/m^4.
$$
Functions $c_i=c_i(k,q,K)$ and $d=d(k,q,K)$ are given by
\begin{eqnarray}
c_1(k,q,K) &= &\frac{M^2(kq)-(Kk)(Kq)}{M^2q^2-(Kq)^2},
\nonumber\\
c_2(k,q,K) &=  &\frac{(Kk)q^2-(Kq)(kq)}{M^2q^2-(Kq)^2}\label{eqn:C}\\
d(k,q,K) &= &\frac{(Kk)^2q^2+M^2(kq)^2+(Kq)^2k^2-M^2q^2k^2
-2(Kq)(Kk)(kq)}{2M^2(M^2q^2-(Kq)^2)}.
\label{eqn:D}
\end{eqnarray}

\newpage
\begin{table}[h]
\caption{\label{tab:1s0} Spin angular momentum parts
$\tilde \Gamma_0^{\tilde\alpha}$
for the $~^1S_0$ channel}
\[
\begin{array}{cc}
\hline\hline
\tilde\alpha&{\sqrt{8\pi} \;\;\tilde \Gamma}_0^{\tilde\alpha}\\[1ex]
\hline
^1S_0&-\gamma_5\\[1ex]
^3P_0&-1/2({\hat k_1}-{\hat k_2})|\bk|^{-1} \gamma_5\\
\hline\hline
\end{array}
\]
\end{table}

\begin{table}[h]
\caption{\label{tab:3s1} Spin angular momentum parts
$\tilde \Gamma_{\cal M}^{\tilde\alpha}$
for $^3S_1-^3D_1$ channel}
\[
\begin{array}{cc}
\hline\hline
\tilde\alpha&{\sqrt{8\pi}\;\;\tilde \Gamma}_{\cal M}^{\tilde\alpha}\\[1ex]
\hline
^3S_1&{\hat \xi_{\cal M}}\\
^3D_1& -\frac{1}{\sqrt{2}}
\left[ {\hat \xi_{\cal M}}+\frac{3}{2}
({\hat k_1}-{\hat k_2})(k\xi_{\cal M})\bk^{-2}\right]\\
^3P_1& \sqrt{\frac{3}{2}}
\left[ \frac{1}{2} {\hat \xi_{\cal M}}({\hat k_1}-{\hat k_2})
-(k\xi_{\cal M}) \right]|\bk|^{-1}\\
^1P_1&\sqrt{3} (k\xi_{\cal M})|\bk|^{-1}\\
\hline\hline
\end{array}
\]
\end{table}

\newpage

\vskip 50mm


\newpage


\vskip 50mm


\begin{figure}[h]

\unitlength=1mm
\special{em:linewidth 0.4pt}
\linethickness{0.4pt}

\hskip 30mm
\begin{picture}(50.00,70.00)
\put(20.00,60.00){\line(1,0){20.00}}
\put(40.00,60.00){\line(3,2){20.00}}
\put(20.00,60.00){\vector(1,0){10}}
\put(40.00,60.00){\vector(3,2){10}}
\put(35.00,40.00){\line(3,2){15.00}}
\put(35.00,40.00){\vector(3,2){7.5}}
\put(50.00,50.00){\line(1,0){14.00}}
\put(50.00,50.00){\vector(1,0){7.5}}
\put(35.00,38.00){\line(1,0){29.00}}
\put(40.00,38.00){\vector(1,0){17.00}}
\put(20.00,40.00){\line(1,0){15.00}}
\put(20.00,38.00){\line(1,0){15.00}}
\bezier{40}(25.00,41.00)(28.00,39.00)(28.00,39.00)
\bezier{40}(25.00,37.00)(28.00,39.00)(28.00,39.00)
\put(35.00,39.00){\circle*{4.00}}
\bezier{40}(40.00,60.00)(42.00,60.00)(42.00,58.00)
\bezier{40}(42.00,58.00)(42.00,56.00)(44.00,56.00)
\bezier{40}(44.00,56.00)(46.00,56.00)(46.00,54.00)
\bezier{40}(46.00,54.00)(46.00,52.00)(48.00,52.00)
\bezier{40}(48.00,52.00)(50.00,52.00)(50.00,50.00)
\put(28.00,30.00){\makebox(0,0)[cc]{D: $K,{\cal M}$}}
\put(73.00,45.00){\makebox(0,0)[cc]{np: $P=K+q$}}
\put(47.00,57.00){\makebox(0,0)[cc]{$q$}}
\put(25.00,63.00){\makebox(0,0)[cc]{$k_e$}}
\put(55.00,74.00){\makebox(0,0)[cc]{$k_e^{\prime}$}}
\end{picture}

\vskip -20mm
\caption{The plane wave approximation.}
\label{fig:pwa}
\end{figure}
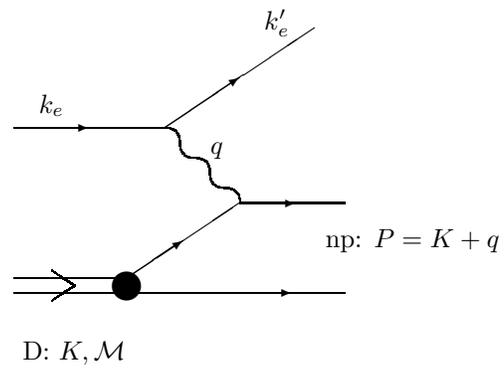

\newpage

\end{document}